# Influence of trapped magnetic field of Sn-Pb solders on electrical resistivity measurement: an example of superconducting transition of Sn

Takumi Ichikawa[1], Yuto Watanabe[1], Takumi Murakami[1], Poonam Rani[1], Aichi Yamashita[1], Yoshikazu Mizuguchi[1]*

[1]*Department of Physics, Tokyo Metropolitan University, 1-1, Minami-osawa, Hachioji, Tokyo 192-0397, Japan*

*E-mail: mizugu@tmu.ac.jp

We examined the affection of the flux-trapped states of Sn60-Pb40 solders on superconducting properties of a Sn wire. The temperature dependence of electrical resistivity at $H = 0$ Oe after zero-field cooling (ZFC) showed a sharp superconducting transition at $T = 3.7$ K. In contrast, that after field cooling (FC) resulted in broadening of the transition. The difference between ZFC and FC data evidences critical affection of trapped fluxes at solders on superconducting states. We propose that, in electrical measurements where magnetic fields of several hundred Oe are critical, field experience should be seriously considered when using solders.

Soldering has been an important technique in several cases; for example, the solders have been used in the production process of electronic circuits and devices and for the fabrication of electric contacts for electric measurements in the field of materials science and condensed matter fields. Since the old years, Sn-Pb solders have been used because of their low melting temperatures [1,2]. Although the use of Sn-Pb solders containing toxic Pb has been avoided in recent years [3], the Sn-Pb solders are still working in some scientific research and industry scenes. Furthermore, solders are used in superconducting joints [4]. Sn-Pb solders are basically phase-separated composite materials with Sn and Pb regions where those elements are separated with a size of several μm [5]. As well known, pure Sn and Pb are superconducting materials with a transition temperature ($T_c$) of 3.7 and 7.2 K, respectively. The Sn-Pb solders trap magnetic fluxes [6,7], and recent analyses of low-temperature physical properties revealed that the Sn regions lose bulkiness of superconductivity due to the trapped magnetic field [5]. In addition, a novel functionality, nonvolatile magneto-thermal switching (MTS), using the magnetic-flux-trapped states in the Sn-Pb solders have been observed [5]. Because nonvolatility of MTS had not been observed in MTS materials [8-11], the use of phase-separated superconducting composites will be critically important for the development of thermal management technologies. The flux-trapped states would, however, at the same time, negatively work in some cases. For example, experimentalists on materials use the solders when measuring electrical resistivity of the target sample, and the samples are sometimes exposed to magnetic fields at low temperatures, even lower than $T_c$ of solder (typically, $T_c$ = 7.2 K). As a fact, our previous work on evaluation of superconducting critical current density of Fe-based superconducting wires also used Sn-Pb solders when fabricating [12]. Since the Fe-based superconductors are robust to magnetic fields, the effects of trapped field on the experiments on the Fe-based wires was negligible, the trapped field exceeding 500 Oe in the Sn-Pb solders [5] should affect several low-temperature experiments. To test the possibility, we measured the temperature dependence of electrical resistivity ($\rho$-$T$) of pure Sn wires with four terminals made by Cu wires and commercial Sn60-Pb40 solder (flux-cored) in several magnetic-field ($H$) conditions (Fig. 1(a)). The zero-field-cooled (ZFC) $\rho$-$T$ data exhibited a conventional sharp superconducting transition at $T$ = 3.7 K. However, the field-cooled (FC) $\rho$-$T$ data exhibited clear broadening of the transition and the lowering of $T_c$. As shown in Figs. 1(b) and 1(c), the difference in $\rho$-$T$ between the ZFC and FC conditions is induced by the partial suppression of the superconducting states of Sn by the trapped field in the FC solders.

We made four terminals for electrical resistivity measurements on Sn wire (0.5 mm in diameter, 3N purity, product of Nilaco) using commercial Sn60-Pb40 (HOZAN) and Cu wire (0.2 mm in diameter). A schematic image of the fabricated four terminals is shown in Fig. 1(a). The $\rho$-$T$ was measured using a Physical Property Measurement System (PPMS Dyna-Cool, Quantum Design) with different magnetic field experiences. We performed the $\rho$-$T$ measurements heating at $H$ = 0 Oe after ZFC and FC under 1500 Oe [FC (1500 Oe)]. In addition, we measured $\rho$-$T$ at $H$ = 100 and -100 Oe after FC (1500 Oe). The magnetic properties of the used Sn60-Pb40 solder were investigated using Magnetic Property Measurement System (MPMS3, Quantum Design) in the SQUID-VSM mode. The $T$ dependence of magnetization ($M$) ($M$-$T$) was measured at $H$ = 10 Oe after ZFC and FC at 10 Oe, respectively. In addition, $M$-$T$ was measured at $H$ = -100, 0, 100 Oe after FC (1500 Oe) to evaluate the $T$ dependence of trapped field. The $M$-$H$ was measured at $T$ = 1.8K at -1500 < $H$ < 1500 Oe.

The results shown here are taken in the same sequence after setting to the target $H$. Figs. 2(a) and 2(b) show the $\rho$-$T$ of Sn wire measured at $H$ = 0 Oe after ZFC and FC (1500 Oe), respectively. For the ZFC data, a sharp superconducting transition is observed at the onset temperature $T_c^{onset}$ = 3.7 K, which is consistent with $T_c$ of pure Sn. The zero-resistivity temperature ($T_c^{zero}$) is close to $T$ = 3.7 K in Fig. 2(a). In Fig. 2(b), the $T_c^{onset}$ is 3.7K but $T_c^{zero}$ is clearly lowered to 2.8 K. In addition, the clear broadening of superconducting transition was observed. Figs. 3(a) and 3(b) show the $\rho$-$T$ of Sn wire measured at $H$ = 100 and -100 Oe after FC (1500 Oe), respectively. Normally, $T_c$ is suppressed by external fields, and $T_c$ does not depend on the direction of the applied magnetic field for polycrystalline materials. In the present experiments, however, magnetic field along one direction is trapped in the solder after FC (1500 Oe). The $T_c^{onset}$ at $H$ = 100 Oe is ~3.3 K as shown in Fig. 3(a). Broadening of superconducting transition is also observed for the data at $H$ = 100 Oe, but the transition is relatively sharper than that at $H$ = 0 Oe both after FC (1500 Oe). As shown in Fig. 3(b), the $T_c^{onset}$ for the data at $H$ = -100 Oe is ~2.8 K. The $T_c^{zero}$ is lower than 1.8 K, and zero-resistivity state was not observed. Furthermore, the superconducting transition was clearly broadened. In supplementary data, the results on the other Sn sample (sample #2) are summarized in Figs. S1 and S2, and reproducibility of this effect has been confirmed on another sample (sample #3) as well. These results clearly demonstrate that the trapped fields have spatial expanse, and the magnetic field trapping in the solders affects electrical measurements, at least a superconducting transition of low-$H_c$ superconductors. Furthermore, if solders are used in a small-scale electronic circuit and experienced magnetic fields, the trapped field at the solders would affect electronic properties of the neighboring components.

To discuss the clearly different $\rho$-$T$ results shown in Figs. 2 and 3, the magnetic properties of the used Sn60-Pb40 solder were studied. Fig. 4(a) shows the $M$-$T$ of Sn60-Pb40 solder measured at $H$ = 10 Oe after ZFC and FC (10 Oe). Here, all the $M$ data are corrected by demagnetization factor. A sharp superconducting transition is observed for both ZFC and FC data at $T \sim 7.2$ K, which is the $T_c$ of pure Pb. As discussed in Ref. 5, the single-step transition without a transition signal at $T_c$ (Sn) = 3.7 K is possibly explained by the μm-size phase separation and proximity effects of superconductivity of the Pb regions. Fig. 4(b) shows the $M$-$T$ of Sn60-Pb40 solder measured at $H$ = 0 Oe after FC (1500 Oe). The data exhibits the $T$ evolution of the trapped field. From the $M$-$T$ data, we estimate the trapped field at $T \sim 3.7$ K is about 250 Oe, which is enough to suppress superconducting states of Sn. Figs. 4(c) and 4(d) show the $M$-$T$ of Sn60-Pb40 solder measured at $H$ = 100 and -100 Oe after FC (1500 Oe), respectively. In Fig. 4(c), trapped field is ~260 Oe at $T$ = 1.8 K, and then the $4\pi M$ decreases with increasing $T$. Furthermore, the $4\pi M$ becomes negative at $T \sim 6.5$ K, reaches a minimum at $T \sim 6.6$ K, and moves up toward 0 G at $T \sim 6.8$ K. In Fig. 4(d), the trapped field is the maximum, ~380 Oe, at $T$ = 1.8 K, and the $4\pi M$ decreases to 0 G at $T \sim 6.8$ K. Noticeably, the trapped field of the Sn60-Pb40 solder becomes higher when a negative (opposite to FC field direction) field is applied after FC. When a positive magnetic field is applied after FC, trapped field is lower. Fig. 4(e) shows the $M$-$H$ of Sn60-Pb40 solder measured at $T$ = 1.8 K at -1500 < $H$ < 1500 Oe, which is totally consistent with the above discussion on the correlation between trapped and applied $H$. The difference in Fig. 2 can be simply understood by the absence and presence of trapped field in ZFC and FC data, respectively. The data shown in Fig. 3 can be understood by considering the difference in the trapped fields. In addition to external field, the solder under $H$ = -100 Oe after FC is trapping a large field exceeding 300 Oe at 3 K. This is the cause of the clear broadening of the transition and the disappearance of zero-resistivity states in Fig. 3(b).

Here, we showed that the Sn-Pb solders with trapped field affect electrical resistivity measurements for superconducting Sn. Among the Sn-Pb solders, Sn-poor composition results in a higher field trapping of ~ 700 Oe [13]. If other phase-separated composites with a higher trapped field, which would be related to $H_c$ or upper critical field of the constituent materials of the composites, the affection of the use of the composites for soldering on the trapped field on electrical measurements, including Hall measurements, cannot be negligible. In addition, the trapped field may affect physical property investigation and application of spin-related materials [14–26] and superconducting joint application [4,27–30].

In conclusion, we measured the temperature dependences of electrical resistivity of a Sn wire as a superconducting sample with four terminals fabricated using Sn60-Pb40 solder and Cu wires. Although ZFC data showed a sharp superconducting transition, FC data exhibited clear broadening. Furthermore, application of negative field after FC (1500 Oe), where the trapped field becomes maximum, zero-resistivity states were not observed at $T >$ 1.8 K. Our present experiments clearly shows that the use of solders, phase-separated superconductors, in low-$T$ magneto-electrical experiments or application of low-$T$ spin/electronic devices would potentially affect physical properties or efficiency of devices.

**Acknowledgments**

The authors thank H. Arima for discussion on magnetic-flux trapping in solders. The work was partly supported by JST-ERATO (JPMJER2201) and TMU Research Project for Emergent Future Society.

# Figures

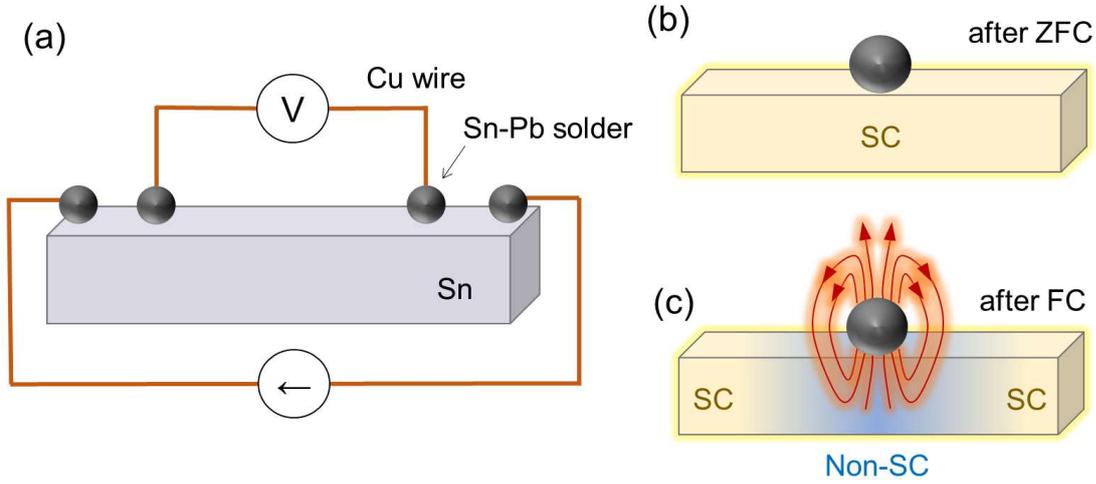

**Fig. 1.** (a) Schematic image of four terminals made on the Sn wire for four-terminal electrical resistivity measurements. (b,c) Absence and presence of the trapped magnetic fields in the Sn-Pb solder after ZFC and FC.

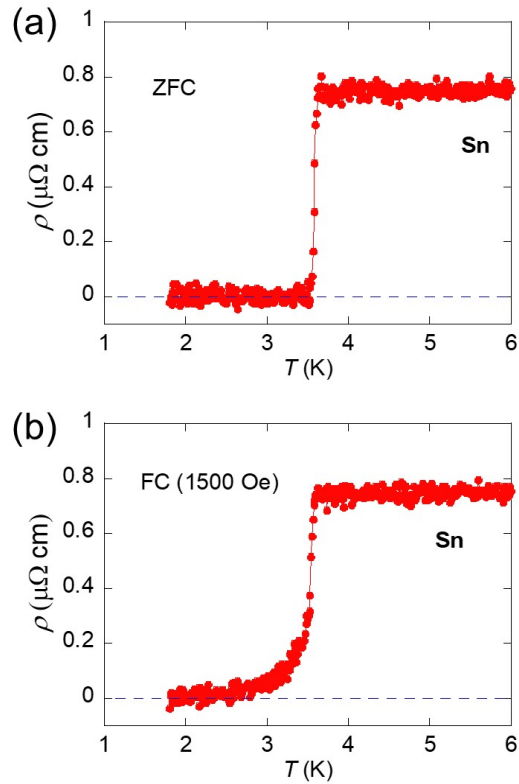

**Fig. 2.** $\rho$-$T$ of the Sn wire measured at $H = 0$ Oe after (a) ZFC and (b) FC at 1500 Oe.

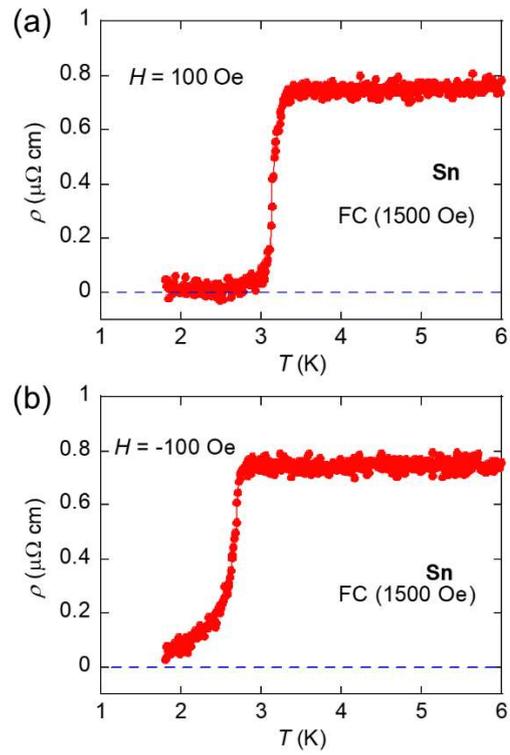

**Fig. 3.** $\rho$-$T$ of the Sn wire measured at (a) $H = 100$ Oe and (b) $H = -100$ Oe after FC at 1500 Oe.

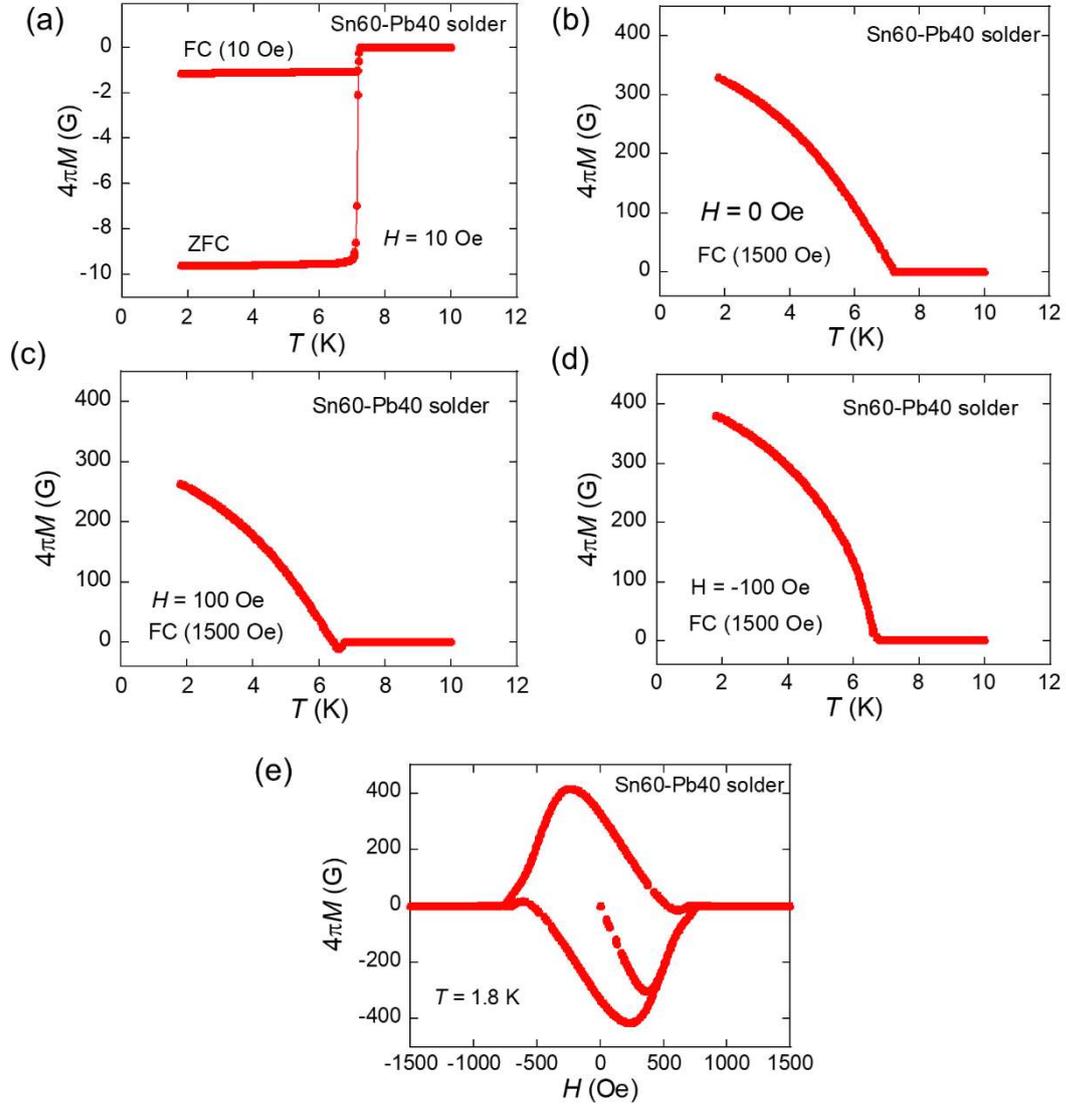

**Fig. 4.** (a) *M-T* of the Sn60-Pb40 solder taken at *H* = 10 Oe after ZFC and FC at 10 Oe. (b) *M-T* of the Sn60-Pb40 solder taken at *H* = 0 Oe after FC at 1500 Oe. (c,d) *M-T* of the Sn60-Pb40 solder taken at (c) *H* = 100 Oe and (d) *H* = -100 Oe after FC at 1500 Oe. (e) *M-H* of the Sn60-Pb40 solder at 1.8 K.

**Supplemental data**

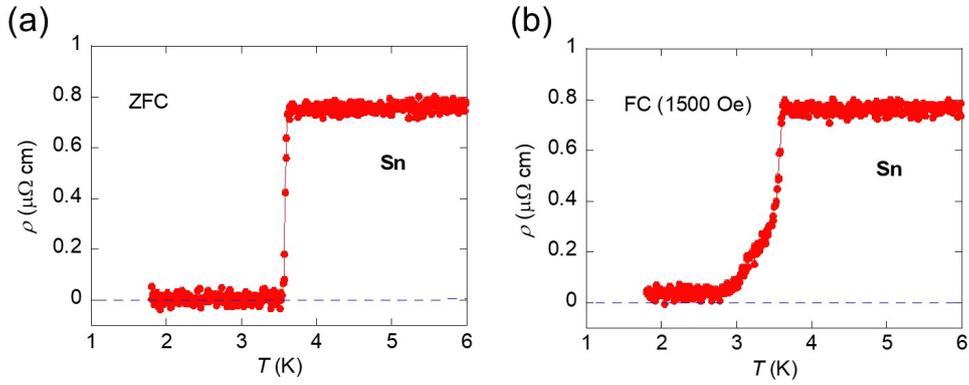

Fig. S1. $\rho$-$T$ of the other Sn wire (sample #2) measured at $H = 0$ Oe after (a) zero-field cooling (ZFC) and (b) field cooling (FC) at 1500 Oe.

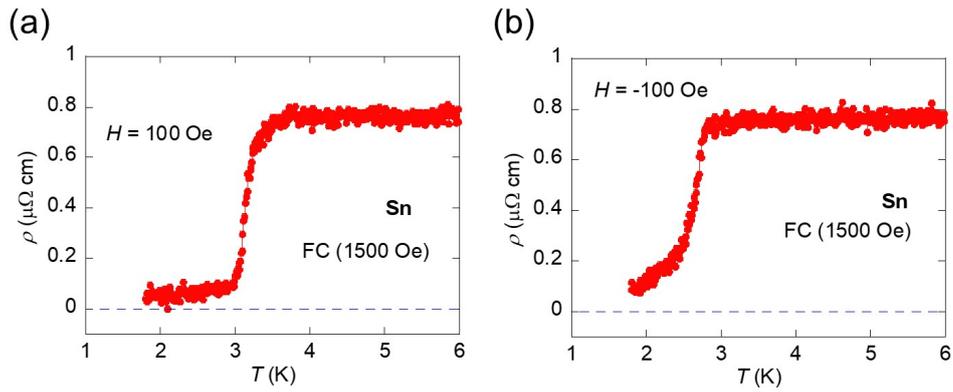

Fig. S2. $\rho$-$T$ of the other Sn wire (sample #2) measured at (a) $H = 100$ Oe and (b) $H = -100$ Oe after FC at 1500 Oe.